\documentstyle[aps]{revtex}
\begin{document}                           
\input{psfig.tex}

\title{Integrability of the critical point of the Kagom\'e 
       three-state Potts model}

\author{J.-Ch. Angl\`es d'Auriac}
\address{Centre de Recherches sur les Tr\`es Basses Temp\'eratures,
B.P. 166, F-38042 Grenoble, France}

\date{\today}

\maketitle

\begin{abstract}
The vicinity of the critical point of the three-state Potts model
on a Kagom\'e lattice is studied by mean of Random Matrix Theory.
Strong evidence that the critical point is integrable is given.
\end{abstract}
\section{Introduction}
The $q$-state Potts model \cite{pot52}, being one of the simplest
generalization of the Ising model, is a challenging model of
classical lattice statistical mechanics. Despite its apparent simplicity
it has not yet been analytically solved for a number of states $q >2$
even in two dimensions (for a review see \cite{wu82}).
The square, the triangular and the hexagonal lattice have
been shown to be integrable {\em at the critical} point.
The free energy the correlation function have been calculated. 
Out of the critical point the free energy has not been calculated, 
moreover strong indications that the model is not 
integrable have been given~\cite{mey}.
On another hand the two dimensional Kagom\'e lattice is of great
interest both from the statistical point of view and from
solid state point of view. 
The anti ferromagnetic Kagom\'e lattice provides an 
example of geometrically frustrated spin system \cite{manip1}.
Several experiments on the so-called SCGO compounds family suggest
the existence of a "spin-liquid" phase transition \cite{manip2}.
It has also been seen
as topological spin glass \cite{premi}.

The critical point of the Potts model on the Kagom\'e
lattice is known in the Ising ($q=2$) case only. However, 
an {\em algebraic} variety has been conjectured to be critical 
for the checkerboard square lattice with multi spin 
interaction in ~\cite{wu79}. From this conjecture the critical point
of the Potts Kagom\'e lattice has been deduced~\cite{wu79}.
This value has recently been confirmed with high precision
by extensive Monte-Carlo simulation~\cite{wu98}, so that
the actual critical point is at least extremely closed
from the conjectured point. 
The aim of this letter is not give another numerical
confirmation of this conjecture, but to
address the question of a possible integrability of the critical
point. Indeed it is possible that the critical point of
the Kagom\'e $q=3$ Potts model is of the same nature as
the critical point of the square, hexagonal or triangular
lattice, ie an integrable point. It is however also possible
that this point is not integrable, as is, for example,
the the critical point of the three dimensional Ising
cubic lattice \cite{mey}. This letter is intended to give an answer to this
question. It is organized as follows: in a first paragraph we sketch
briefly the basics of Random Matrix Theory (RMT) applied to classical 
lattice statistical mechanics, and recall how it acts as an 
`integrability detector'. In the next paragraph we give a more 
convenient formulation of the Kagom\'e lattice as a
checkerboard Interaction-Round-a-Face (IRF) model, and
we discuss the symmetries of the corresponding transfer
matrix. In the final paragraph the results are presented
and discussed.

\section{The RMT methodology}
The RMT lattice statistical mechanics has first been introduced
for Quantum one dimensional system \cite{tou,hsu},
it has then been extended to classical two dimensional lattices \cite{37}.
Roughly speaking the main idea is to compare the given Hamiltonian
to an `average Hamiltonian'. Among all Hamiltonians,
integrable Hamiltonians are extremely peculiar and show
significant difference with `average Hamiltonian'.
To be more specific one is interested in the statistical
properties of the discrete spectrum viewed as an infinite
set of real numbers. One compares the statistical
properties of the spectrum of the system under consideration
with the one of random matrices from a suitable statistical ensemble.
For an integrable system it is found that the spectrum as many
properties of a set of {\em independent} numbers, ie the suitable
matrix statistical ensemble is the set of the diagonal matrices
with independent normal centered entries. By contrast, for a
non-integrable time reversal symmetric system the spectrum is
well described by spectrum of matrices from the so-called
Gaussian Orthogonal Ensemble (GOE) of symmetric matrices with independent
normal centered entries. The eigenvalue spacing
distribution $P(s)$ discriminates well between these two limiting
cases: $P(s) \, ds$ is the probability that the difference between
two consecutive eigenvalues belongs to the interval $[s,s+ds]$.
It is simple to show that for independent eigenvalues one has
$P(s)=e^{-s}$. For GOE matrices $P(s)$ is extremely
close (actually strictly equal for $2 \times 2$ matrices) to
the Wigner surmise $W(s)={\pi \over 2} s \exp(-\pi s^2/4)$. 
Provided that the proper normalization
explained below has been performed, and if the system size
is large enough, then the distribution $P(s)$ of the system
under consideration is close to one of the exponential or Wigner
distribution. To quantify this ``proximity'' one introduces
the parametrized distribution
\begin{equation}
P_{\beta}(s)=c(1+\beta)s^{\beta} \exp\left(-cs^{\beta+1} \right)\;,
\label{bb}
\end{equation}
with $c=\left[ \Gamma \left({\beta+2 \over \beta+1} \right) \right]^{1+\beta}$,
which is one of the infinitely many possible interpolations between 
the exponential
and Wigner distribution. For $\beta=0$ one recovers the exponential law
and for $\beta=1$ one recovers the Wigner law. The full lines
on Fig.~\ref{pds} represents these two distributions. Parameter
$\beta$ is interpreted as the repulsion between eigenvalues.
$P(s)$ is a quantity involving only pairs of eigenvalues;
a quantity involving more than two eigenvalues
is the spectral rigidity
\begin{equation}
\Delta_3(\lambda)=
\left\langle {1 \over \lambda} \min_{a,b} \int_{\alpha-l/2}^{\alpha+\lambda/2}
(N_u(\epsilon)-a\epsilon-b)^2 d\epsilon \right\rangle_{\alpha}\;,
\label{e:rigidity}
\end{equation}
where $N_u(\epsilon) \equiv \sum_i \Theta(\epsilon-\epsilon_i)$ is the
integrated density of unfolded eigenvalues and $\langle \dots \rangle_{\alpha}$
denotes an average over $\alpha$. It provides a finer test
of the closeness of a spectrum
with a GOE spectrum or with independent numbers.
These two limiting cases are shown of Fig.~\ref{D3} as full lines.
For the classical lattice system, like spin models or vertex models, 
the {\em transfer matrix}~\cite{42} is considered instead of the 
Hamiltonian which generally has a trivial spectrum.

The operational procedure has been explained in details
in~\cite{41,42}. It consists in two main steps. One first 
sort the states according to their symmetries: states of
different symmetries have to be considered separately. Then,
since universal properties cannot be found on the row spectrum,
a procedure of ``normalization'', known as the unfolding,
is performed: it consists in imposing a {\it local} density
of eigenvalues of the order of unity. Then the
statistical properties of the sorted and unfolded spectrum 
are analyzed. Note that this analysis requires real eigenvalues.
This is generally insured by the  of the Hamiltonian.

\section{Transfer matrix of the Potts model on the Kagom\'e lattice}
\begin{figure}
\psfig{file=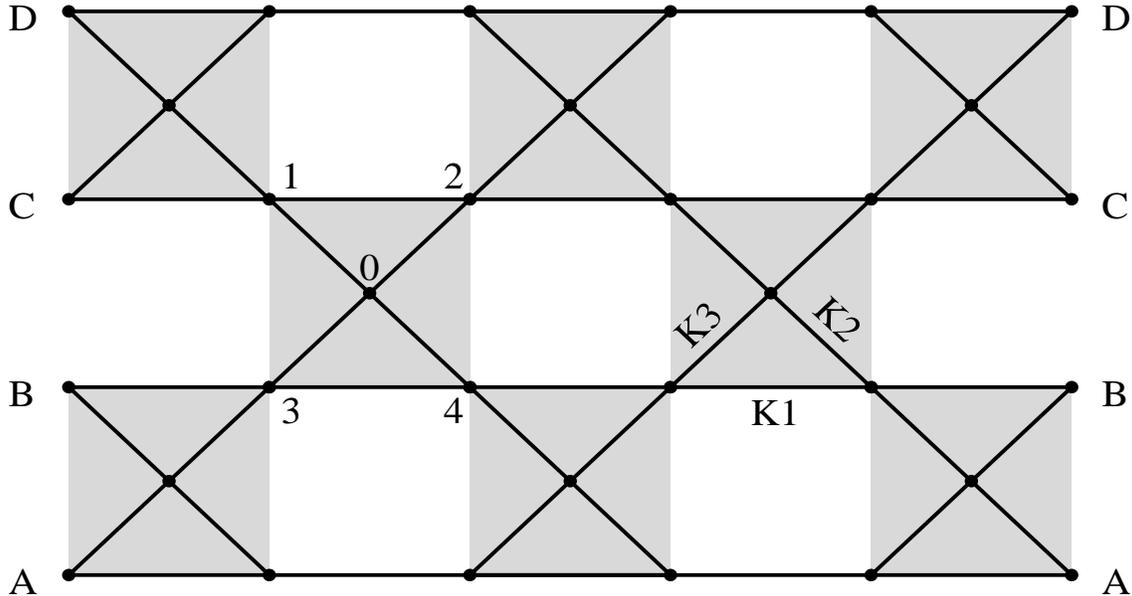,height=10cm,width=16cm}
\caption{Correspondence between the Kagom\'e lattice and the
checkerboard IRF model.
\label{corresp}
}
\end{figure}

The Kagom\'e lattice is a two dimensional lattice
depicted on Figure \ref{corresp}. The black circles
represent the sites, and the lines are the coupling bonds.
The partition function reads
\begin{equation}
\label{hamiltonien}
Z_{\rm kag} = \prod_{<i,j>}{K_{i\,j}^{\delta(\sigma_i , \sigma_j)}}
\end{equation}
$<i,j>$ denotes the neighboring sites on the Kagom\'e lattice,
$\delta$ is the Kronecker symbol,
$K_{i\,j} = K_1, K_2$ or $K_3$ for each of the three types
of bonds, and $\sigma_i$ can take one of the three values 0, 1, 2.
The Kagom\'e lattice can be regarded as a lattice of
basic ``cells'' (1,2,3,4) as shown on Fig.~\ref{corresp}. 
Each cell consists in a up pointing
triangle sharing a site (site labelled 0 on the figure)
with a down pointing triangle.
These cells are disposed on every other (say black) plaquettes
of checkerboard. To see the correspondence with an
IRF model on a checkerboard lattice, one perform a partial summation on the
central site (site 0), seeking coupling constants such that
\begin{equation}
W_{\rm IRF}(\sigma_1,\sigma_2,\sigma_3,\sigma_4) = 
\sum_{\sigma_0} W_{\rm kag}(\sigma_1,\sigma_2,\sigma_3,\sigma_4;\sigma_0)
\end{equation}
This is possible if one introduces multi spin interaction. The
general correspondence is given in Appendix~\ref{appa}.
For the isotropic $K_1=K_2=K_3=K$ case the Boltzmann weights
of the plaquettes are given by
\begin{equation}
W_{\rm kag}(\sigma_1,\sigma_2,\sigma_3,\sigma_4;\sigma_0) = 
K^{\delta_{0,1}+\delta_{0,2}+\delta_{0,3}+\delta_{0,4}+
   \delta_{1,2}+\delta_{3,4}}
\end{equation}
and
\begin{equation}
W_{\rm IRF}(\sigma_1,\sigma_2,\sigma_3,\sigma_4) = 
L_h^{\delta_{1,2} + \delta_{3,4}} \,
L_v^{\delta_{1,3} + \delta_{2,4}} \,
L_d^{\delta_{1,4} + \delta_{2,3}} \,
T^{\delta_{1,2,3} + \delta_{1,2,4}
 + \delta_{1,3,4} + \delta_{2,3,4}} \,
F^{\delta_{1,2,3,4}}
\label{ww}
\end{equation}
where $\delta_{i,j}=\delta(\sigma_i,\sigma_j)$
and $\delta_{i,j,k}=\delta(\sigma_i,\sigma_j,\sigma_k)$ 
($\sigma_i=1,2,3,4$).
The parameters of the IRF model are given as a function of
the parameters of the Kagom\'e by:
\begin{eqnarray}
L_h&=&K L_v=K L_d=\frac{2 K^2+1}{K+2} \nonumber\\
T&=&\frac{(K+2)(K^3+K+1)K}{(2 K^2+1)^2}\\
F&=& \frac{(K^4+2)(2 K^2+K+1)^3}{(K^3+K+1)^4} \nonumber 
\end{eqnarray}

From a computational point of view it is more convenient to use
the IRF representation, this is the form chosen in the calculations
presented here. A first difficulty in constructing
the transfer matrix comes from the fact the actual
transfer matrix which brings from row A to row C on Fig.~\ref{corresp}
is the product of the two transfer matrices bringing
respectively from row A to row B and from row B to row C.
As a result the calculation of a single entry of the transfer
matrix involves a combination of sums and products over
$N$ terms, $N = 3^L$ being the size of the matrix ($L$ is
the linear size of the rows, $L=6$ on Fig.~\ref{corresp}).
This greatly slows down the calculations. Another difficulty
comes with the fact that, unlike simpler lattice, there is no 
way to make symmetric the transfer matrix $T$, so that
the eigenvalues are complex. It is still possible to deal
with complex eigenvalues. In this case integrability also
leads to {\em independent} eigenvalues, and non integrability
leads to repulsion. However, in practice, it is much 
more difficult to make the distinction, since in two dimensions
repulsion is less effective. Therefore we have chosen to analyze
the statistical properties of the operator
${\cal T}=(T + \tilde{T})/2$.

In the actual computation, the transfer matrix calculated
is the diagonal-to-diagonal transfer matrix. As explained
in the previous paragraph one needs to find the symmetries of the
operator ${\cal T}$. Let us define the translation operator $t$,
the reflection $r$, and the color operator $c$
\begin{eqnarray}
t \; | \sigma_0,  \sigma_1, \cdots ,\sigma_{L-1} >  &=&  
         | \sigma_1, \sigma_2,  \cdots \sigma_0 > \\
r \; | \sigma_0,  \sigma_1, \cdots ,\sigma_{L-1} >  &=&  
         | \sigma_{L-1}, \sigma_{L-2},  \cdots ,\sigma_0 > \\
c \; | \sigma_0,  \sigma_1, \cdots ,\sigma_{L-1} >  &=&  
         | \sigma_0+1, \sigma_1+1,  \cdots ,\sigma_{L-1}+1 > 
\end{eqnarray}
where the sum are taken modulo 3. It is easy to see the three operators
$t$, $r$ and $c$ commute with ${\cal T}$, and that $c$ commutes
with both $t$ and $r$. Thus the symmetry
group is the direct product ${\cal D}_L \otimes{\cal D}_3$ 
where  ${\cal D}_L$ is
the dihedral group of index $L$. The irreducible representations of this
group are simply the matrix product of elements of ${\cal D}_L$
and $ {\cal D}_3$ (see details in \cite{45}). 
The sizes of the blocks corresponding to the different
representations are given in Tab.~\ref{tabsize}. In the statistical
analysis the blocks of size smaller than 100 have been discarded 
to minimize small size effects. 

\section{Results and Conclusion}
\begin{figure}
\psfig{file=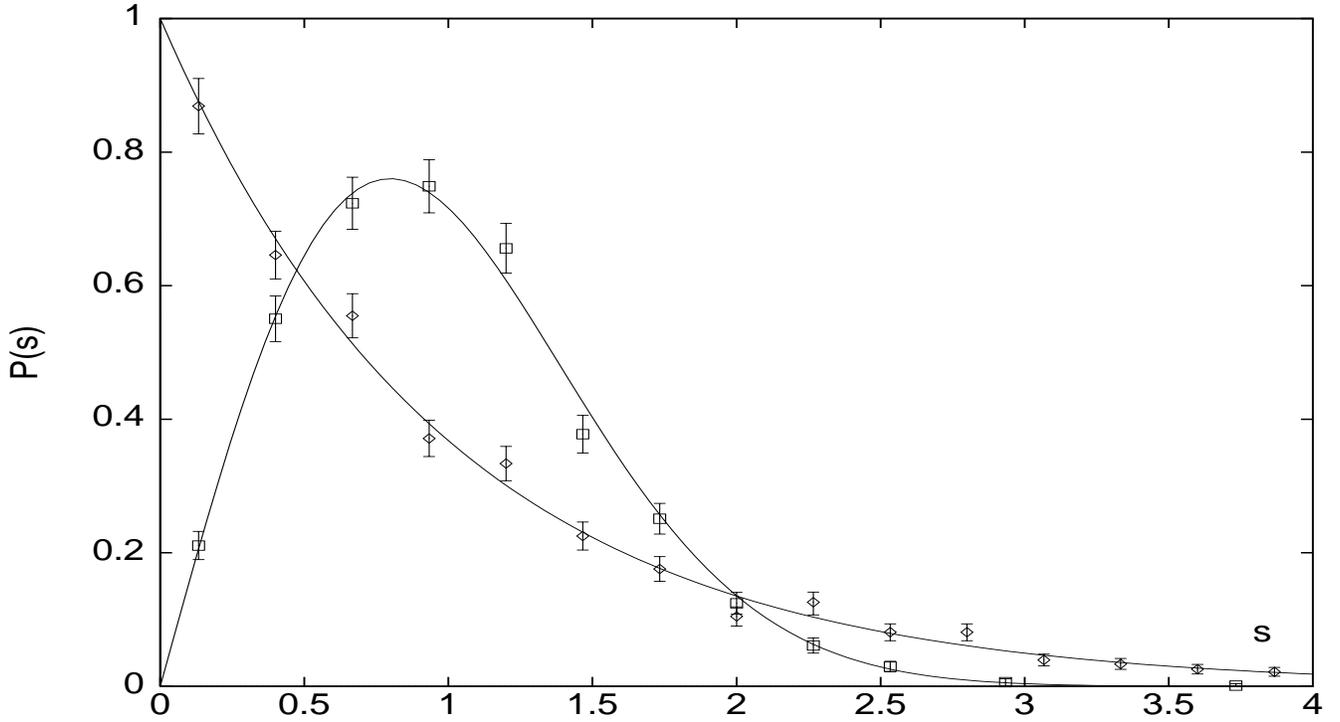,height=10cm,width=18cm}
\caption{Distribution of eigenvalue spacings for $K \simeq K_c$.
The linear size of the lattice is $L=8$, leading to Hilbert
space of size 6561. 1910 eigenvalues are included in the statistic.
The diamonds corresponds to the critical point $K=K_c$,
and the square corresponds to the the value $K=2$.
\label{pds}
}
\end{figure}

\begin{figure}
\psfig{file=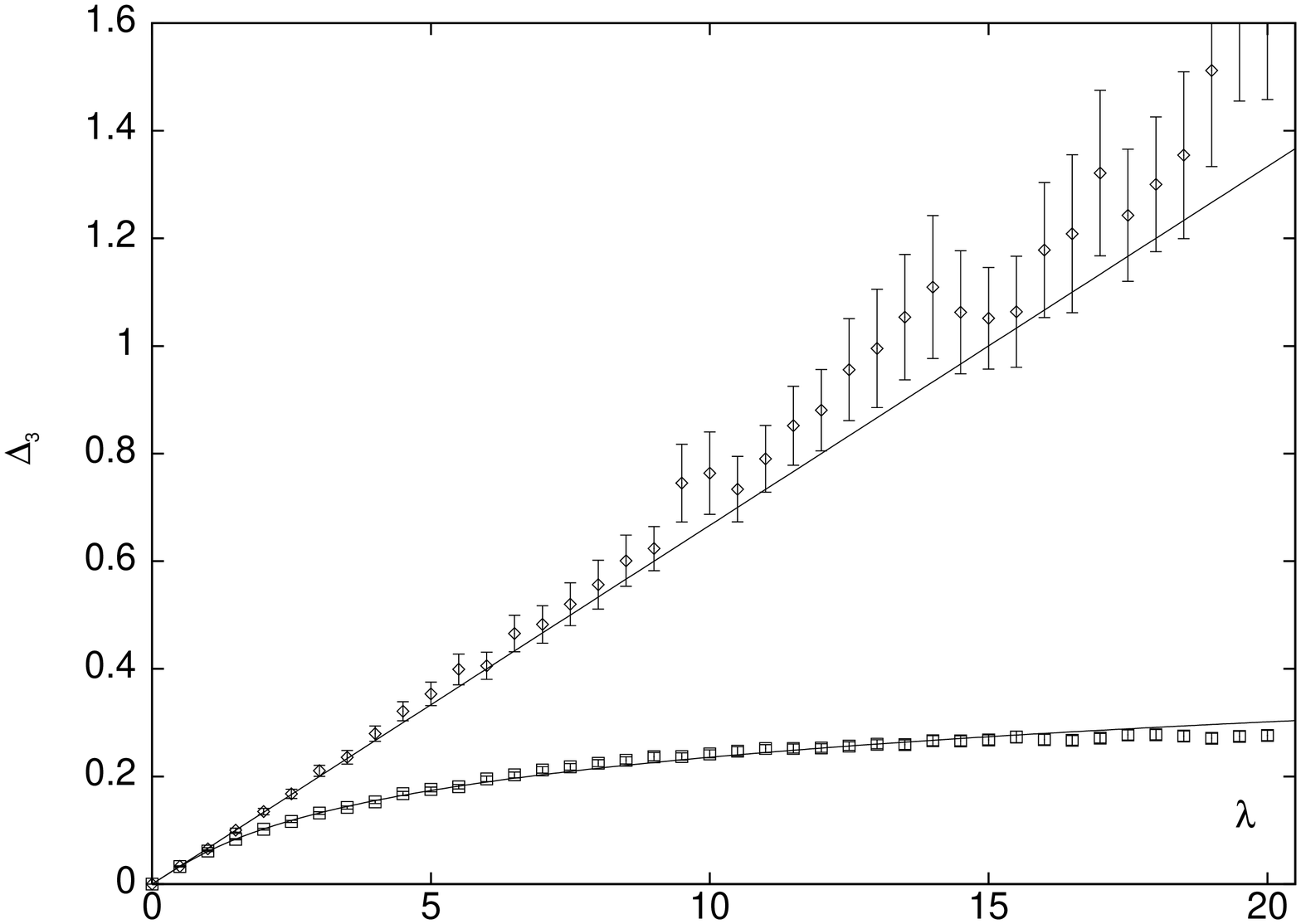,height=10cm,width=18cm}
\caption{Rigidity $\Delta_3$ as a function of $\lambda$ for the same
data as in Fig.~\ref{pds}.
\label{D3}
}
\end{figure}

Using the IRF representation of the Kagom\'e lattice 
we have calculated for different values
of the exponential of the coupling constant $K$ the
symmetrized diagonal-to-diagonal transfer matrix ${\cal T}(K)$ for $L=8$.
The size of the matrix ${\cal T}$ is $3^8=6561$. 
For sizes smaller than $L=8$,
the blocks are quite small, giving poor statistics. On another hand
the size $L=9$ is out of reach of your computing possibility, especially due
to the facts that ${\cal T}$ is a matrix {\em product} 
with all entries  non zero
(in contrast with most quantum Hamiltonian) as explained
in the previous paragraph. So we stick to the case $L=8$ 
and perform the analysis for different values of $K$. 
The matrix is projected in the different invariant 
subspaces, yielding matrix blocks of smaller
sizes (see Tab.~\ref{tabsize}). 
Within blocks there is no degeneracy left, proving that the symmetry group
used is the largest symmetry group for this model, as it should.
The spectrum of the 13 larger blocks is then unfolded,
leading to 2187 eigenvalues distributed in 13 sub-sets.
The eigenvalues spacing distribution $P(s)$ as well as the
rigidity $\Delta_3$ are calculated. 
We present on Fig.~\ref{pds} and Fig.~\ref{D3}
the results for two values of $K$. Firstly for $K=K_c$ where $K_c$
is the value predicted by the Wu's conjecture, which is known to be at 
least extremely close the the actual critical value. And secondly 
for $K=2$ which is sufficiently far from the critical value, and neither too
small nor too big: extremal values leads to accuracy problems.
The results are unambiguous: close to the transition point the
distribution of eigenvalues is very close to an exponential, whereas
out of this point it is close to a Wigner distribution. This is clearly
seen on Fig.~\ref{pds} where the numerical data are presented for both
cases together with the exponential law and the Wigner distribution.
The best fit values of $\beta$ (see Eq.~\ref{bb})
are $\beta(K_c)=0.01$ and $\beta(2)=1.02$. This is a strong indication that
the critical point is integrable. To further support this result, the
rigidity Eq.~\ref{e:rigidity} has also been calculated for the same
value of $K$. The results are shown on Fig.~\ref{D3}, where the numerical
data and the two limiting cases of a GOE spectrum and of independent numbers
are presented. The agreement between the expected behavior and the observed
behavior is very good, at least up to an ``energy scale'' of the order of 15,
(ie 15 eigenvalues since the mean density of eigenvalues is one).
The number variance $\Sigma^2(l)=\left\langle \left[N_u(\epsilon + l/2)
-N_u(\epsilon - l/2)-l \right]^2 \right\rangle_{\epsilon}$ 
where the brackets denote an averaging over $\epsilon$ has also 
been calculated,
giving results not presented here, in perfect agreement with the results 
for the spectral rigidity $\Delta_3$.

In conclusion a RMT analysis performed on the symmetrized transfer matrix
of the Kagom\'e lattice seen as an IRF checkerboard model shows that the
critical point, very close or equal to the Wu's conjecture, is an integrable
point. This is in agreement the conformal theories of the
two dimensional lattice statistical mechanics.
The only example of a critical point where the transfer matrix
eigenvalues spacing distribution is of Wigner type is the 3-dimensional
Ising case.

\begin{table}
\begin{tabular}{|c|c|c|c|c|c|c|c|c|c|c|c|c|c|c|c|c|c|c|c|c|}\hline
0  &1 &2  &3 &4 &5  &6 &7 &8  &9 &10&11 &12 &13 &14 &15 &16 &17 &18 &19 &20 \\\hline
1  &1 &2  &1 &1 &2  &1 &1 &2  &1 &1 &2  &2  &2  &4  &2  &2  &4  &2  &2  &4  \\ \hline
100&66&166&46&66&112&72&53&125&72&80&152&130&140&270&142&134&276&130&140&270\\ 
\end{tabular}
\label{tabsize}
\caption{Sizes and degeneracies of the different blocks for $L=8$.
The first row is the label of the representation, the second is the dimension
of the representation (ie degeneracy of the corresponding eigenvalues)
and the third row is the size of the block to be diagonalized.}
\end{table}

\acknowledgments{
I acknowledge many discussions with J.M. Maillard
during the completion of this work, and with F.Y. Wu during a meeting 
held in Grenoble in September 1998 in F. Jaegger's Memoriam.
I acknowledge also R. Melin for discussions concerning experimental
realizations of the Kagom\'e lattice.}


\appendix

\section{Kagom\'e lattice and checkerboard IRF model}
\label{appa}
For a non-isotropic Kagom\'e lattice the Boltzmann weight
of a basic cell (see Fig~\ref{corresp}) is
\begin{equation}
W_{\rm kag}(\sigma_1,\sigma_2,\sigma_3,\sigma_4;\sigma_0) = 
K_1^{\delta_{1,2}+\delta_{3,4}}
K_2^{\delta_{0,1}+\delta_{0,4}}
K_3^{\delta_{0,2}+\delta_{0,3}}
\end{equation}
and for IRF checkerboard it is
\begin{equation}
W_{\rm IRF}(\sigma_1,\sigma_2,\sigma_3,\sigma_4) = 
L_{12}^{\delta_{1,2}} L_{34}^{\delta_{3,4}}
L_{13}^{\delta_{1,3}} L_{24}^{\delta_{2,4}} 
L_{14}^{\delta_{1,4}} L_{23}^{\delta_{2,3}} 
L_{123}^{\delta_{1,2,3}} L_{124}^{\delta_{1,2,4}}
L_{134}^{\delta_{1,3,4}} L_{234}^{\delta_{2,3,4}} 
L_{1234}^{\delta_{1,2,3,4}}
\left( L_{12}^{34} \right) ^{\delta_{12} \delta_{34}}
\left( L_{13}^{24} \right) ^{\delta_{13} \delta_{24}}
\end{equation}
The notation are the same as in the text Eq.~\ref{ww}.
The parameters of the IRF model are given as a function of
the parameters of the Kagom\'e by:
\begin{eqnarray}
L_{12}(K_1,K_2,K_3) &=& L_{34}(K_1,K_2,K_3) = K_1 L_{13}(K_2,K_3) = K_1 L_{24}(K_2,K_3)
 \nonumber \\
&=&K_1 \frac{(K_2 K_3 + K_2 + K_3) (1 + K_2^2 + K_3^2)}
	      {2 K_2^3 + K_2^2 K_3^2 + 4 K_2 K_3 + 2 K_3^3} \nonumber\\
L_{14}(K_2,K_3)&=&L_{23}(K_3,K_2) \nonumber \\
&=& \frac{K_2^2+K_3^2+1}{2 K_2+2 K_3^2} \nonumber \\
L_{234}(K_2,K_3)&=&L_{134}(K_3,K_2)=L_{124}(K_3,K_2)=L_{123}(K_2,K_3) \nonumber \\
&=&\frac{(K_3^2+2 K_2)(K_2 K_3^2 + K_2+1)(K_2^2+2K_3)^2}
{((K_2 K_3 + K_2 + K_3)(K_2^2 +K_3^2 +1))^2} \nonumber \\
L_{12}^{34}(K_2,K_3)&=&L_{13}^{24}(K_2,K_3) \nonumber \\
&=& \frac{(2 K_2 K_3 +1)(2 K_2^3+K_2^2 K_3^2 + 4 K_2 K_3 + 2 K_3^3)}
         {(K_2^2 +K_3^2 +1)(K_2 K_3 +K_2 +K_3)^2} \nonumber \\
L_{1234}(K2,K3) &=& \frac{(K_2^2 K_3^2+2)(K_2^2+K_3^2+1)^4}
           {((K_2+2 K_3)(K_3+2 K_2))^2} \nonumber \\
& &\times \frac{(K_2 K_3+K_3+K_2)^4}
           {((K_2^2 K_3 + K_3 +1)(K_2 K_3^2 +K_2+1))^2} \nonumber
\end{eqnarray}


\begin{references}

\bibitem{pot52}
R.B. Potts\\
Proc. Camb. Phil. Soc. {\bf 48},106 (1952)

\bibitem{wu82}
F.Y. Wu\\
Rev. Mod. Phy. {\bf 54},235 (1982)

\bibitem{mey}
H. Meyer. \\
Ph.D. dissertation, Grenoble (1996).

\bibitem{manip1}
Y.J. Uemera {\em et al.}\\
Phys. Rev. Lett.{\bf 73}, 3306 (1994).

\bibitem{manip2}
P. Schiffer and I. Daruka,\\
Phys. Rev. {\bf b56}, 13712 (1997).

\bibitem{premi}
P. Chandra, P. Coleman and I. Ritchey\\
J. Phys. I France {\bf 3} 591 (1991).

\bibitem{wu79} F.Y. Wu.\\
	J. Phys. {\bf C12}, L645 (1979).

\bibitem{wu98} 
J.A. Chen, C.K. Hu and F.Y. Wu.\\
	J. Phys. {\bf A},  (1998).

\bibitem{tou}
G. Montambaux, D. Poilblanc, J Bellissard  and C. Sire.\\
Phys. Rev. Lett. {\bf 70} 497, (1993).

\bibitem{hsu}
T. Hsu, J.C. Angl\`es d'Auriac.\\
Phys. Rev. {\bf B47} 21 14291 (1993).

\bibitem{37}
H. Meyer and J. C. Angl\`es d'Auriac.\\
J. Phys. A: Math. Gen. {\bf 29} L483 (1996).

\bibitem{41}
H. Meyer, J. C. Angl\`es d'Auriac and J.M. Maillard.\\
Phys. Rev. {\bf E55}, 5380 (1997).

\bibitem{42}
H. Meyer, J. C. Angl\`es d'Auriac and H. Bruus\\
Phys. Rev. {\bf E55}, 6608 (1997).

\bibitem{45}
H. Bruus and J. C. Angl\`es d'Auriac.\\
Phys. Rev. {\bf B55}, 9142 (1997).



\end{references}
\end{document}